\documentclass[prd,aps,nofootinbib,floatfix,11pt]{revtex4}
\usepackage{amsmath,graphicx,epsfig,amssymb,dsfont,mathtools,}
\usepackage[usenames]{color}
\usepackage{ulem} 
\usepackage{bigstrut}
\usepackage{slashed}
\usepackage{multirow}
\usepackage{subfigure}

\allowdisplaybreaks

\begin{document}\title{Searching for doubly charmed baryon from $\overline{B}_c$ meson decays}

\author{Ye Xing}
\email{Corresponding author. xingye_guang@cumt.edu.cn}
\affiliation{School of Physics, China University of Mining and Technology, Xuzhou 221000, China}

\author{Ji Xu}
\email{Corresponding author. xuji_phy@zzu.edu.cn}
\affiliation{School of Nuclear Science and Technology, Lanzhou University, Lanzhou 730000, China}
\affiliation{School of Physics and Microelectronics, Zhengzhou University, Zhengzhou, Henan 450001, China}

\begin{abstract}
 In this paper, we study the production of doubly charmed baryon from anti-bottom charmed meson. Using the effective Lagrangian approach, we discuss triangle diagrams in hadronic level to get access to the branching ratios of $\overline{B}_c\to \mathcal{B}_{ccq}+\mathcal{B}_{\bar c\bar q\bar q}$. It seems that the specific process $\overline {B}_c \to \Xi_{cc}^{+} \, \overline {\Xi}_{\bar c}^{'0}$ occupies the largest possibility in the order of $9.5\times 10^{-5}$. In addition, although the production of undiscovered $\Omega_{cc}^+$ is Cabibbo suppressed in $\overline B_c\to \Omega_{cc}^+ \, \overline {\Xi}_{\bar c}^0$, it's branching ratio can still reach $10^{-6}$ level. These results are excepted to be fairly valuable supports for future experiments.
\end{abstract}
\maketitle


\section{phenomenological analysis}
The LHCb collaboration claimed the experimental evidence of doubly charmed baryon candidate $\Xi_{cc}^{++}$ in the $\Lambda_c^+ K^- \pi^+\pi^+$ final state in 2017\,\cite{LHCb:2017iph}. Its mass was determined to be $3621.55$ MeV~\cite{LHCb:2019epo}, and its lifetime measured by the weakly decays, was $0.256$ ps\,\cite{LHCb:2018zpl}. Prior to this, the doubly charmed baryon $\Xi_{cc}^+$  with a mass of $3518.7$ MeV was first reported by the SELEX collaboration in decay modes $\Xi_{cc}^+\to \Lambda_c^+ K^- \pi^+$ and $\Xi_{cc}^+ \to p\, D^+ K^-$\,\cite{SELEX:2002wqn,SELEX:2004lln}, its lifetime was found to be less than 33 fs. However, subsequent experiments FOCUS\,\cite{Ratti:2003ez}, BaBar\,\cite{BaBar:2006bab}, Belle\,\cite{Belle:2006edu} and LHCb\,\cite{LHCb:2021eaf} have not confirmed this state yet. Likewise, for $\Omega_{cc}^+$ baryon, the most recent searches from LHCb collaboration still have not reported significant detection signal. Therefore more theoretical and experimental efforts propelling the study of doubly heavy baryons are urgently called for.

It is of key importance for completing the hadron spectrum and revealing the nature of perturbative and non-perturbative QCD dynamics about doubly heavy baryons. Presently various theoretical studies, such as discussions carried by lattice QCD\,\cite{Briceno:2012wt,Brown:2014ena,Perez-Rubio:2015zqb,Padmanath:2013zfa,Alexandrou:2017xwd}, quark model\,\cite{Xing:2021enr,Zhao:2023yuk,Ebert:2002ig,Liu:2019vtx,Chen:2016iyi,Xiao:2017udy,Shah:2017jkr,Yin:2019bxe}, QCD sum rules\,\cite{Wang:2010hs,Azizi:2014jxa,Aliev:2012ru,Shi:2020qde,Shi:2019fph,Wang:2017mqp}, heavy baryon chiral perturbation theory\,\cite{Meng:2022ozq,Liu:2018euh,Li:2017pxa,Shanahan:2014cga} have been applied into the research area of doubly heavy system. For ground states of $\Xi_{cc}$, the majority theoretical predictions concentrate on masses and lifetimes, i.e., the masses are declared in the range from 3.5-3.7 GeV, being close to current experimental measurements. Due to effect of the destructive Pauli interference of the $c$ quark decay products and the valence $u$ quark in the initial state, the lifetime of $\Xi_{cc}^{++}$ is expected to be around 2-4 times larger than the one of $\Xi_{cc}^+$, consequently most of theoretical predictions show that the lifetime of $\Xi_{cc}^+$ is in the range of 40-160 fs. For ground state $\Omega_{cc}^+$, there are also plentiful theoretical predictions, for instance the mass is predicted to be around 3.6-3.9 GeV, and its lifetime is 75-180 fs. We arrange some typical results into Tab.\,\ref{tab:mass}.

The aim of our work is to present a new study about production of doubly charmed baryon from $\overline B_c$ meson, it may be achieved by LHCb and future b-factory.  We derive the production processes in terms of effective Lagrangian method\,\cite{Wise:1992hn,Casalbuoni:1996pg}. The complete transition is divided into weakly decay part associated with transition $\mathcal{O}^i:\ \bar b\to\bar c c\bar d/\bar s$, and strongly couple part related to two strong coupled vertices, i.e., $\mathcal{B}_{cc}\mathcal{B}_cD$ and $\mathcal{B}_{c}\mathcal{B}_{c}J/\psi$ ($\mathcal{B}$ is the general baryon), the production can then be described by some triangle diagrams formally in hadronic level.
Under the SU(3) light quark flavor symmetry\,\cite{Wang:2017azm,Shi:2017dto,Wang:2018utj,Xing:2018bqt,Xing:2022aij,Li:2023kcl,Han:2023teq}, it is convenient to relate different production channels, wherein the final states can be one doubly charmed baryon $\mathcal{B}_{cc}$ ($\Xi_{cc}^{++}$, $\Xi_{cc}^+$, $\Omega_{cc}^+$), and anti-charmed triplet baryon $\mathcal{B}_{\bar c3}$ or anti-sextet baryon $\mathcal{B}_{\bar c\bar 6}$, 
\begin{eqnarray}
&&\Gamma(\overline B_c \to \Xi_{cc}^{++} \overline \Xi_{\bar c}^{-})=\Gamma(\overline B_c \to \Xi_{cc}^+ \, \overline \Xi_{\bar c}^0) \,,\quad
\Gamma(\overline B_c \to \Xi_{cc}^{++} \overline \Lambda_{\bar c}^{-})=\Gamma(\overline B_c \to \Omega_{cc}^{+} \, \overline \Xi_{\bar c}^{0}) \,,\nonumber\\
&&\Gamma(\overline B_c \to \Xi_{cc}^{++} \overline \Xi_{\bar c}^{'-})=\Gamma(\overline B_c \to \Xi_{cc}^+ \, \overline \Xi_{\bar c}^{'0}) \,,\quad
\Gamma(\overline B_c \to \Xi_{cc}^{++} \overline \Sigma_{\bar c}^{-})=2 \Gamma(\overline B_c \to \Xi_{cc}^{+} \, \overline \Sigma_{\bar c}^0) \,.
\end{eqnarray}
Furthermore, the rates of decay widths are related to the Cabibbo-Kobayashi-Maskawa (CKM) matrix elements $V_{cs}$ and $V_{cd}$,
\begin{eqnarray}\label{symmetryanalysis}
\frac{\Gamma(\overline B_c \to \Xi_{cc}^{++} \overline \Xi_{\bar c}^{-})}{\Gamma(\overline B_c \to \Xi_{cc}^{++} \overline \Lambda_{\bar c}^{-})}=\frac{\Gamma(\overline B_c \to \Xi_{cc}^{++} \overline \Xi_{\bar c}^{'-})}{\Gamma(\overline B_c \to \Xi_{cc}^{++} \overline \Sigma_{\bar c}^{-})}=\frac{|V_{cs}|^2}{|V_{cd}|^2} \,,
\end{eqnarray}
Consistently, we calculate the branching ratio from effective Lagrangian method, comparing with the predictions from SU(3) analysis in the hope of providing assistance to understand and search for doubly heavy baryons.
\begin{table}
\caption{The masses and lifetimes of doubly charmed baryon predicted from several theoretical methods or experiments.}\label{tab:mass}
\begin{tabular}{|c|c|c|c|c|c c| c|}
  \hline
 &LQCD\cite{Brown:2014ena} & QM\cite{Ebert:2002ig} & [1]  & QCDSR\cite{Azizi:2014jxa} &\multicolumn{2}{c|}{Exp(mass \& lifetime)}\\\hline
  $\Xi^{++}_{cc}$ & 3.610\,GeV & 3.676\,GeV & 3.627\,GeV & 3.72\,GeV & 3.622\,GeV\,\cite{LHCb:2019epo} & 0.256 ps\,\cite{LHCb:2018zpl}\\
   $\Xi_{cc}^+$ & 3.610\,GeV & 3.676\,GeV & 3.627\,GeV & 3.72\,GeV &3.519\,GeV\,\cite{SELEX:2004lln}& $<33$ fs~\cite{SELEX:2004lln}\\
   $\Omega^+_{cc}$ & 3.738\,GeV & 3.832\,GeV & $-$ &3.73\,GeV &$-$&$-$\\
  \hline
\end{tabular}
\end{table}

The rest of this paper is organized as follows. In Sec.\,\ref{Thebranchingratios}, we simply introduce the calculation framework about the production of doubly charmed baryons from $\overline B_c$ decay, several amplitudes of triangle diagrams according to the effective Lagrangian method are achieved. Then we give numerical analysis and branching ratios of different processes in Sec.\,\ref{Numericalanalysis}. The last section contains a brief summary.


\begin{figure}
\includegraphics[width=0.98\columnwidth]{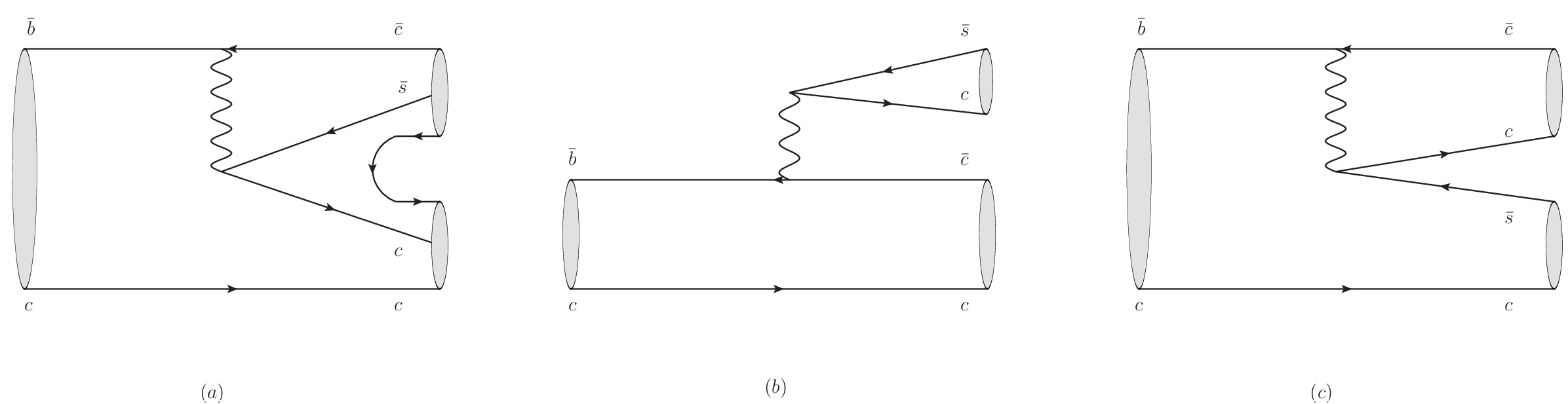}
\caption{(a). The production processes of $\overline B_c\to \mathcal{B}_{cc}+\overline {\mathcal{B}}_{\bar c}$, which are non-factorizable and color suppressed. (b,c). The weak decays of $\overline B_c \to D(D^*)+J/\psi(\eta_c)$, which possess the factorizable external and internal $W$-emission diagrams respectively.}
\label{fig:weakdecay}
\end{figure}
\renewcommand\thesubsection{(\roman{subsection})}

\section{The branching ratios}
\label{Thebranchingratios}
\begin{figure}
\includegraphics[width=0.75\columnwidth]{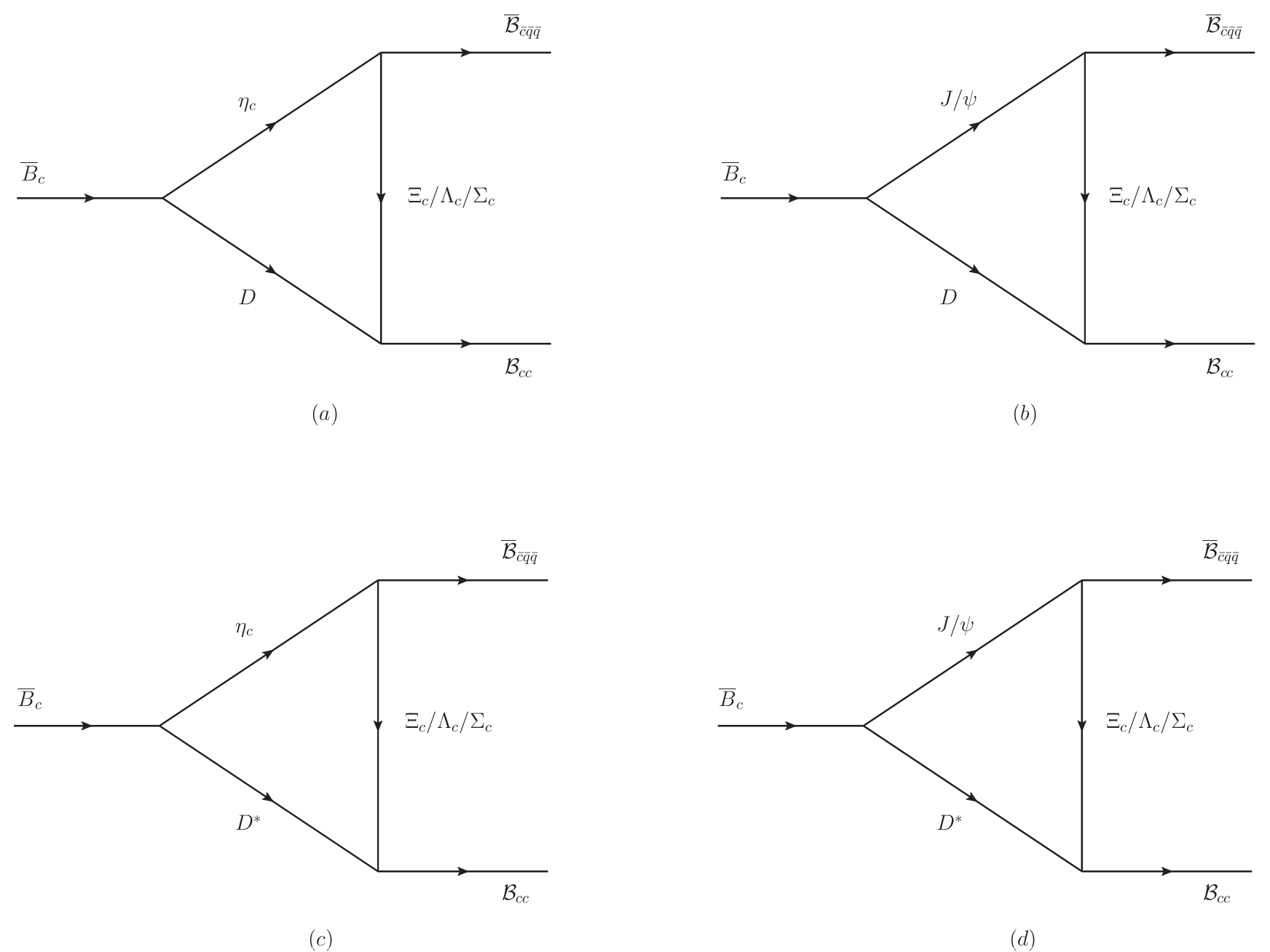}
\caption{The triangle diagrams attached to the production of doubly charmed baryons, brought by exchanging charm baryons $\Xi_c,\Lambda_c,\Sigma_c$.}
\label{fig:production}
\end{figure}

The production of doubly charmed baryons $\mathcal{B}_{cc}$ from $\overline B_c$ occurs through the weak decay of bottom quark $\bar b \to \bar c c \bar d/\bar s$, whose topological diagram in quark level is shown in Fig.\,\ref{fig:weakdecay}(a). The non-factorizable color suppressed diagram with two baryons in final states are unusual, it is hard to handle directly by common factorization schemes. In this paper, we shall consider the process within the framework of the effective Lagrangian approach---a convenient method to study hadrons in hadronic level. The total production matrix element can be further divided into weakly transition matrix as well as strongly coupled matrix by inserting some complete basis,
\begin{eqnarray*}
\langle\mathcal{B}_{cc} \overline {\mathcal{B}}_{\bar c}|\mathcal{H}_{eff}|\overline B_c\rangle=\sum_{\lambda}\langle\mathcal{B}_{cc} \overline {\mathcal{B}}_{\bar c}|\mathcal{H}_{\lambda}|\lambda\rangle\langle\lambda|\mathcal{H}_{eff}|\overline B_c\rangle.
\end{eqnarray*}
It is not hard to see that the dominant contribution respecting with weakly transition matrix corresponds to $\langle D(D^*)\eta_c|\mathcal{H}_{eff}|\overline B_c\rangle$ or $\langle D(D^*)J/\psi|\mathcal{H}_{eff}|\overline B_c\rangle$ (here, we only consider the major contribution $\lambda=D(D^*)J/\psi(\eta_c)$), which is depicted with an accessible emission diagram\,\cite{Colangelo:2003sa} in Fig.\,\ref{fig:weakdecay}(b,c). Meanwhile, the remaining one can be decoded by several effective Lagrangians which will be given below.
Combining them, the productions of spin-$\frac{1}{2}$ doubly charmed baryons $\mathcal{B}_{cc}$ ($\Xi_{cc}^{++},\Xi_{cc}^+,\Omega_{cc}^+$) stemming from the triangle diagrams can be attained. We proceed with processes under quark transition $\bar b\to \bar c c\bar s$:
\begin{eqnarray*}
\overline B_{c}\to \Xi_{cc}^{++}+ \overline \Xi_{\bar c}^- (/\Xi^{'-}_{\bar c})\,, \qquad \overline B_{c}\to \Xi_{cc}^{+}+\overline \Xi_{\bar c}^0 (/\Xi^{'0}_{\bar c}) \,,
\end{eqnarray*}
and the transition $\bar b\to \bar c c \bar d$:
\begin{eqnarray*}
\overline B_{c}\to \Xi_{cc}^{++}+\overline \Lambda_{\bar c}^- (/\overline \Sigma_{\bar c}^-)\,, \qquad \overline B_{c}\to \Xi_{cc}^{+}+\overline \Sigma_{\bar c}^0 \,, \qquad \overline B_{c}\to\Omega_{cc}^+ +\overline \Xi_{\bar c}^0 \,.
\end{eqnarray*}
The corresponding triangle diagrams are drawn in Fig.\,\ref{fig:production}. Here, we have excluded some unallowable processes in phase space, as well as the processes with decoupling vertex.

\subsection{Effective Lagrangians}
We adopt the effective Lagrangian to study the production of doubly charmed baryons. The process can be regarded as $\overline B_c$ meson firstly decays into $D(D^*)$ and $J/\psi\,(\eta_c)$ mesons, then exchanging with charm baryon $\mathcal{B}_c$ to produce anti-charmed baryon $\overline{\mathcal{B}}_{\bar c}$ and doubly charmed baryon $\mathcal{B}_{cc}$, $\overline B_c \to M_1M_2 \xrightarrow[]{\mathcal{B}_c} \overline{\mathcal{B}}_{\bar c}\mathcal{B}_{cc}$ ($M_1M_2$ can be $J/\psi \,D, \eta_c \,D, J/\psi \,D^*$ or $\eta \,D^*$). The weak decay $\overline B_c \to J/\psi \,(\eta_c)+ D \,(D^*)$ can occur via the factorizable $W$-emission process which is shown in Fig.\,\ref{fig:weakdecay}(b), it usually offers the largest contribution according to the topological classification of weak decays. The effective Hamiltonian is
\begin{eqnarray}
\mathcal{H}_{eff} &=& \frac{G_F}{\sqrt{2}} V_{cb}^*V_{cs} (C_1(\mu) \mathcal{O}_1(\mu)+C_2(\mu) \mathcal{O}_2(\mu))+h.c. \,,
\end{eqnarray}
here, $G_F$ is the Fermi constant, $C_{1,2}(\mu)$s are the Wilson coefficients, and $\mathcal{O}_{1,2}(\mu)$s are the tree level fermion operators, with $\mathcal{O}_{1}=(\bar b_{\alpha} c_{\beta})_{V-A}(\bar c_{\beta}s_{\alpha})_{V-A}$ and $\mathcal{O}_{2}=(\bar b_{\alpha} c_{\alpha})_{V-A}(\bar c_{\beta}s_{\beta})_{V-A}$. The combinations of the effective Wilson coefficient $a_{1,2}$ appears as $a_1=C_1+C_2/N_c$, $a_2=C_2+C_1/N_c$ with $N_c$ the number of colors.

The amplitude of the weak process can be expressed as the products of two current hadronic matrix elements companied by form factor and decay constant,
\begin{eqnarray}
&\mathcal{M}(\overline B_c \to J/\psi D(D^*))&=\frac{G_F}{\sqrt{2}} V_{cb}^* V_{cs} a_1  \langle J/\psi|(\bar b c)_{V-A} |\overline B_c\rangle \langle D(D^*)|(\bar cs)_{V-A} |0\rangle\nonumber\\ &&+\frac{G_F}{\sqrt{2}} V_{cb}^* V_{cs} a_2  \langle J/\psi|(\bar b c)_{V-A} |\overline B_c\rangle \langle D(D^*)|(\bar cs)_{V-A} |0\rangle\,,\\
&\mathcal{M}(\overline B_c \to \eta_c D(D^*))&=\frac{G_F}{\sqrt{2}} V_{cb}^* V_{cs} a_1 \langle \eta_c|(\bar b c)_{V-A} |\overline B_c\rangle \langle D(D^*)|(\bar cs)_{V-A} |0\rangle\nonumber\\
&&+\frac{G_F}{\sqrt{2}} V_{cb}^* V_{cs} a_2 \langle \eta_c|(\bar b c)_{V-A} |\overline B_c\rangle \langle D(D^*)|(\bar cs)_{V-A} |0\rangle \,.
\end{eqnarray}
The matrix elements between pseudoscalar or vector meson and vacuum have the following formations\,\cite{Cheng:1996cs},
\begin{eqnarray}
\langle P|(\bar c s)_{V-A} |0\rangle=i f_{P} \, p_{\mu} \,, \qquad \langle V| (\bar cs)_{V-A}|0 \rangle=m f_{V} \varepsilon^*_{\mu} \,.
\end{eqnarray}
where P, V stand for pseudoscalar and vector mesons; $f_{P}$ and $f_{V}$ are the pseudoscalar and vector meson decay constants; $\varepsilon^*_{\mu}$ represents the polarization vector of vector meson.

We follow the parameterized form of the transition $\overline B_c \to J/\psi D \,(D^*)$\,\cite{Dhir:2008hh},
\begin{eqnarray}\label{eq:formfactor}
\langle J/\psi(P_V)|(\bar b c)_{V-A} |\overline B_c(P_{B_c})\rangle\ &=&\frac{2}{m_{B_c}+m_V}\varepsilon_{\mu\nu\rho\sigma}\varepsilon^{*\nu}P_{B_c}^{\rho}
P_V^{\sigma} V(q^2)+i\varepsilon_{\mu}^*(m_{B_c}+m_V)A_1(q^2)\nonumber\\\nonumber
&&-i\frac{\varepsilon^*\cdot q}{m_{B_c}+m_V}(P_{B_c}+P_{V})_{\mu} A_2(q^2)-i \frac{\varepsilon^*\cdot q}{q^2} 2 m_V q_{\mu} A_3(q^2)\\
&&+i \frac{\varepsilon^*\cdot q}{q^2} 2m_V q_{\mu} A_0(q^2) \,,\\
\langle \eta(P_P)|(\bar b c)_{V-A} |\overline B_c(P_{B_c})\rangle\ &=&(P_{B_c}+P_P-\frac{m_{B_c}^2-m_P^2}{q^2}q)_{\mu} F_1(q^2)+\frac{m_{B_c}^2-m_P^2}{q^2}q_{\mu} F_0(q^2) \,.
\end{eqnarray}
here $\varepsilon_{\mu}$ denotes the polarization vector of $J/\psi$, transition momentum $q_{\mu}\equiv(P_{B_c}-P_P)_{\mu}$. In order to cancel the poles at $q^2=0$, invariant weak form factors $F_0(q^2),\, F_1(q^2),\, A_0(q^2)$ and $A_3(q^2)$ satisfy the following conditions:
\begin{eqnarray}
F_1(0)=F_0(0)\,,\quad A_3(0)=A_0(0)\,,\quad A_3(q^2)=\frac{m_{B_c}+m_V}{2m_V}A_1(q^2)-\frac{m_{B_c}-m_V}{2m_V}A_2(q^2) \,.\nonumber
\end{eqnarray}
In the SU(4) symmetry, the interaction Lagrangians related to the processes are given as\,\cite{Yalikun:2021dpk}
\begin{eqnarray}
\mathcal{L}_{\mathcal{B}\mathcal{B}P} &=& \frac{g_{\mathcal{B}\mathcal{B}P}}{m_{P}}{} \langle\overline{\mathcal{B}} \gamma_5\gamma_{\mu} \partial^{\mu}P \mathcal{B}\rangle \,,\\
\mathcal{L}_{\mathcal{B}\mathcal{B}V} &=& g_{\mathcal{B}\mathcal{B}V}{} \langle\overline{\mathcal{B}} \gamma_{\mu} V^{\mu} \mathcal{B}\rangle+ \frac{f_{\mathcal{B}\mathcal{B}V}}{2 m_V} \langle \overline{\mathcal{B}} \sigma_{\mu\nu} \partial_{\mu}V^{\nu} \mathcal{B} \rangle \,.
\end{eqnarray}
Where $g_{\mathcal{B}\mathcal{B}P}$ and $g_{\mathcal{B}\mathcal{B}V}$ are the couplings of pseudoscalar meson or vector meson and two baryons $\mathcal{B}\mathcal{B}$. Since there is no experimental data to extract values of the needed couplings, we deduce them from the generic SU(4) symmetry, i.e., the empirical values $g_{NN\pi}=13.5,\, g_{\rho NN}=3.25,\, f_{\rho NN}=\kappa_{\rho} g_{\rho NN},\, \kappa_{\rho}=6.1$\,\cite{Machleidt:1987hj,Liu:2001ce}, are used to reach the couplings between doubly charmed baryon and charmed baryon. One can resort to Appendix\,\ref{sec:appenda} for more details. Within SU(4) symmetry, pseudoscalar and vector mesons reclassified as 15-plet, whose expressions are provided as\,\cite{Lin:1999ve}
\begin{eqnarray}
P=\frac{1}{\sqrt{2}}\left(
\begin{array}{cccc}
  \frac{\pi^0}{\sqrt{2}}+\frac{\eta}{\sqrt{6}}+\frac{\eta_c}{\sqrt{12}} & \pi^+ & K^+ & \overline{D}^0 \\
  \pi^- & -\frac{\pi^0}{\sqrt{2}}+\frac{\eta}{\sqrt{6}}+\frac{\eta_c}{\sqrt{12}} & K^0 & D^- \\
  K^- & \overline K^0 & -\sqrt{\frac{2}{3}}\eta+\frac{\eta_c}{\sqrt{12}} & D_s^-\\
  D^0& D^+ &D_s^+ & -\frac{3\eta_c}{\sqrt{12}}
\end{array}
\right).
\end{eqnarray}
and
\begin{eqnarray}
V_{\mu}=\frac{1}{\sqrt{2}}\left(
\begin{array}{cccc}
\frac{\rho^0}{\sqrt{2}}+\frac{\omega}{\sqrt{6}}+\frac{J/\psi}{\sqrt{12}} & \rho^+ & K^{*+} & \overline D^{*0}\\
\rho^- & -\frac{\rho^0}{\sqrt{2}}+\frac{\omega}{\sqrt{6}}+\frac{J/\psi}{\sqrt{12}} & K^{*0} & D^{*-} \\
K^{*-} & \overline{K}^{*0} & -\sqrt{\frac{2}{3}}\omega+\frac{J/\psi}{\sqrt{12}} & D_s^{*-}\\
D^{*0} & D^{*+} & D_s^{*+} &- \frac{3 J/\psi}{\sqrt{12}}
\end{array}
\right).
\end{eqnarray}
Baryons respesented as $\mathcal{B}$ belong to a 20-plet,
\begin{eqnarray}
&& \mathcal{B}^{121} = p, \ \mathcal {B}^{122} = n, \  \mathcal{B}^{132} = \frac{1}{\sqrt {2}}\Sigma^0 - \frac{1}{\sqrt{6}}\Lambda, \ \mathcal{B}^{213} = \sqrt{\frac{2}{3}}\Lambda, \nonumber\\
&& \mathcal{B}^{231} = \frac{1}{\sqrt{2}}\Sigma^0 + \frac{1}{\sqrt {6}}\Lambda, \ \mathcal {B}^{232} = \Sigma^-, \ \mathcal{B}^{233} = \Xi^-, \ \mathcal {B}^{311} = \Sigma^+, \nonumber\\
&& \mathcal {B}^{313} = \Xi^0, \ \mathcal {B}^{141} = -\Sigma_c^{++}, \ \mathcal {B}^{142} = \frac {1} {\sqrt {2}}\Sigma_c^++\frac {1}
{\sqrt {6}}\Lambda_c, \ \mathcal {B}^{143} = \frac{1}{\sqrt {2}}{\Xi'}_c^+ -\frac{1}{\sqrt {6}}\Xi_c^+, \nonumber\\
&& \mathcal {B}^{241} = \frac {1} {\sqrt {2}}\Sigma_c^+-\frac{1}{\sqrt {6}}\Lambda_c, \ \mathcal {B}^{242} = \Sigma_c^0, \  \mathcal{B}^{243} = \frac {1} {\sqrt {2}} {\Xi'} _c^0 + \frac {1} {\sqrt {6}}\Xi_c^0, \ \mathcal {B}^{341} = \frac {1} {\sqrt {2}} {\Xi'}_c^++\frac {1} {\sqrt {6}}\Xi_c^+, \nonumber\\
&& \mathcal {B}^{342} = \frac {1} {\sqrt {2}} {\Xi'} _c^0 - \frac {1}{\sqrt {6}}\Xi_c^0, \ \mathcal {B}^{343} = \Omega_c^0, \ \mathcal{B}^{124} = \sqrt {\frac {2} {3}}\Lambda_c, \ \mathcal {B}^{234} = \sqrt {\frac {2} {3}}\Xi_ {c}^0, \nonumber\\
&& \mathcal {B}^{314} = \sqrt {\frac {2} {3}}\Xi_c^+, \ \mathcal{B}^{144} = \Xi_ {cc}^{++}, \ \mathcal {B}^{244} = -\Xi_ {cc}^+, \ \mathcal {B}^{344} = \Omega_ {cc}^+.
\end{eqnarray}
These states are expressed by tensors $\mathcal{B}^{ijk}$, where the first two indices is anti-symmetric.
\subsection{Amplitudes and Decay Rates}
Having the effective Lagrangians at hand, we can obtain the decay amplitudes of Fig.\,\ref{fig:production}. For instance, the amplitude of $\overline B_c\to \Xi_{cc}^{++}\overline \Xi_{\bar c}^-$ is
\begin{eqnarray*}
M_a \!\!&=&\!\! \int \frac{d^4k_1}{(2\pi)^4}g_{\text{\tiny $\Xi_{cc}\Xi_c D$}}g_{\text{\tiny $\overline \Xi_{\bar c}\Xi_c \eta_c$}} \frac{\bar u_{\text{\tiny $\Xi_{cc}$}}(p_2)\gamma_{\nu}(/\!\!\!k_1-m_{\text{\tiny $\Xi_{c}$}})\gamma_{\mu}\nu_{\text{\tiny $\overline \Xi_{\bar c}$}}(p_1)k_3^{\mu}k_2^{\nu} }{(k_1^2-m_{\text{\tiny $\Xi_{c}$}}^2)\ (k_2^2-m_D^2)\ (k_3^2-m_{\eta_c}^2)} \mathcal{F}^2(k_1^2)
(i f_D)\\&&\Big((p\cdot k_2+k_3\cdot k_2-m_{B_c}^2+m_{\eta}^2)F_1(k_2^2)+(m_{B_c}^2-m_{\eta}^2)F_0(k_2^2)\Big) \,,\\
M_b \!\!&=&\!\! i\int \frac{d^4k_1}{(2\pi)^4} g_{\text{\tiny $\Xi_{cc}\Xi_c D$}}g_{\text{\tiny $\overline \Xi_{\bar c}\Xi_c \psi$}} \frac{\bar u_{\text{\tiny $\Xi_{cc}$}}(p_2)(/\!\!\!k_1-m_{\text{\tiny $\Xi_{c}$}})\gamma_5(\gamma_{\mu}-i \kappa_{\psi}\sigma_{\alpha \mu}k_3^{\alpha}/(2m_{\psi}))\nu_{\text{\tiny $\overline \Xi_{\bar c}$}}(p_1)}{(k_1^2-m_{\text{\tiny $\Xi_{c}$}}^2)\ (k_2^2-m_D^2)\ (k_3^2-m_{\Psi}^2)} \mathcal{F}^2(k_1^2) (i f_D)ik_{2\nu}\\&&(-g^{\mu\nu}+\frac{k_3^{\mu} k_3^{\nu}}{m_{\psi}^2})
 \Big((m_{B_c}+m_{\psi})A_1(k_2^2)-\frac{(p+k_3)\cdot k_2}{m_{B_c}+m_{\psi}}A_2(k_2^2)-2m_{\psi} A_3(k_2^2)+2m_{\psi} A_0(k_2^2)\Big) \,,\\
M_c \!\!&=&\!\! i\int \frac{d^4k_1}{(2\pi)^4} g_{\text{\tiny $\Xi_{cc}\Xi_c D^*$}}g_{\text{\tiny $\overline \Xi_{\bar c}\Xi_c \eta_c$}} \frac{\bar u_{\text{\tiny $\Xi_{cc}$}}(p_2)(\gamma_{\mu}-i \kappa_{D^*}\sigma_{\alpha \mu}k_2^{\alpha}/(2m_{D^*}))(/\!\!\!k_1+m_{cqq})\gamma_5(/\!\!\!k_1+/\!\!\!p_1)\nu_{\text{\tiny $\overline \Xi_{\bar c}$}}(p_1)}{-(k_1^2-m_{\text{\tiny $\Xi_{c}$}}^2)\ (k_2^2-m_{D^*}^{2})\ (k_3^2-m_{\eta_c}^2)}\\&&\mathcal{F}^2(k_1^2) (-g^{\mu\nu}+\frac{k_2^{\mu} k_2^{\nu}}{m_{D^*}^2})
 (m_{D^*}f_{D^*})\Big((p+k_3-\frac{(m_{B_c}^2-m_{\eta}^2)}{k_2^2}k_2)^{\nu}F_1(k_2^2) +\frac{(m_{B_c}^2-m_{\eta}^2)}{k_2^2} k_2^{\nu} F_0(k_2^2)\Big) \,,\\
M_d \!\!&=&\!\! \int \frac{d^4k_1}{(2\pi)^4} \frac{\bar u_{\text{\tiny $\Xi_{cc}$}}(p_2)(\gamma_{\mu}-i\kappa_{D^*} \sigma_{\alpha' \mu}k_2^{\alpha'}/(2m_{D^*}))(/\!\!\!k_1+m_{\text{\tiny $\Xi_{c}$}})(\gamma_{\nu}-i \kappa_{\psi} \sigma_{\beta' \nu}k_3^{\beta'}/(2m_{\psi}))\nu_{\text{\tiny $\overline \Xi_{\bar c}$}}(p_1)}{-(k_1^2-m_{\text{\tiny $\Xi_{c}$}}^2)\ (k_2^2-m_{D^*}^2)\ (k_3^2-m_{\Psi}^2)} \\&&(g_{\text{\tiny $\Xi_{cc}\Xi_c D^*$}}g_{\text{\tiny $\overline \Xi_{\bar c}\Xi_c \psi$}})\mathcal{F}^2(k_1^2) (m_{D^*}f_{D^*})(-g^{\mu\alpha}+\frac{k_2^{\mu} k_2^{\alpha}}{m_{D^*}^2})
(-g^{\nu\beta}+\frac{k_3^{\nu} k_3^{\beta}}{m_{\psi}^2})\Big(\frac{2}{m_{B_c}+m_{\psi}}\varepsilon_{\alpha\beta\rho\sigma}p^{\rho}k_2^{\sigma}V(k_2^2)\\&&+i g_{\alpha\beta}(m_{B_c}+m_{\psi})A_1(k_2^2)-i\frac{(p+k_3)_{\alpha}k_{2\beta}}{m_{B_c}+m_{\psi}}A_2(k_2^2)-i \frac{k_{2\alpha}k_{2\beta}}{k_2^2} 2m_{\psi} A_3(k_2^2)+ i\frac{k_{2\beta}}{k_2^2} 2m_{\psi} k_{2\alpha} A_0(k_2^2)\Big) \,.
\end{eqnarray*}
where $k_3=p_1+k_1$ is the momentum of $J/\psi \,(\eta_c)$ and $k_2=p_2-k_1$ is the momentum of $D\,(D^*)$ respectively. Hadrons have finite sizes, therefor the monopole form factor is adopted at each vertex,
\begin{eqnarray*}
\mathcal{F}(k_1^2)=\Big(\frac{\Lambda^2-m_i^2}{\Lambda^2-k_1^2}\Big)^2 \,,
\end{eqnarray*}
here $\Lambda$ is the so-called cutoff mass, which governs the range of suppression. We take the value $\Lambda=m_{E}+\alpha$\,\cite{Xing:2022aij}, $m_E$ is the mass of exchanged particle. In this work, we take naive value for $\alpha$ in range of 100-300 MeV.
The decay rate for nonleptonic transition $\overline{B}_c\to P_1P_2$ is expressed in terms of the decay amplitude $\mathcal{M}(\overline{B}_c\to P_1P_2)$
\begin{eqnarray*}
\Gamma(\overline{B}_c\to \mathcal{B}_{cc}+\overline {\mathcal{B}}_{cqq})=\frac{|\mathbf{P_1}|}{8\pi m_{B_c}^2}|\mathcal{M}|^2 \,,
\end{eqnarray*}
where $\mathbf{P_1}$ is the magnitude of three momentum of the final state meson, given as $|\mathbf{P_1}|=\frac{1}{2m_{B_c}}\sqrt{\lambda(m_{B_c}^2,m^2_{\mathcal{B}_{cc}},m^2_{\overline {\mathcal{B}}_{\bar c}})}$; $\lambda(a,b,c)=a^2+b^2+c^2-2ab-2bc-2ac$.

\section{Numerical analysis}
\label{Numericalanalysis}
The calculation of branching ratios requires the decay constants of several pseudoscalar and vector mesons, which are presented as follows.
\begin{eqnarray*}
f_{D}=0.208\ \text{GeV} \,,\quad f_{D_s}=0.273\ \text{GeV} \,,\quad f_{D^*}=0.245\ \text{GeV} \,,\quad f_{D^*_s}=0.273\ \text{GeV} \,.
\end{eqnarray*}
Furthermore, the masses of mesons and baryons refer to PDG\,\cite{Workman:2022ynf}. Numerous strong couplings are deduced from SU(4) symmetry shown in Appendix\,\ref{sec:appenda}.

On the physical region, each of the transition form factors can be accurately interpolated by the following function\,\cite{Yao:2021pyf}:
\begin{eqnarray*}
f(q^2)=\alpha_1 +\alpha_2 q^2+\frac{\alpha_3 q^4}{m_{B_c}^2-q^2} \,,
\end{eqnarray*}
where the parameters $\alpha_{1,2,3}$ are taken from Tab.\,\ref{tab:formfactor}.

\begin{figure}
\includegraphics[width=0.9\columnwidth]{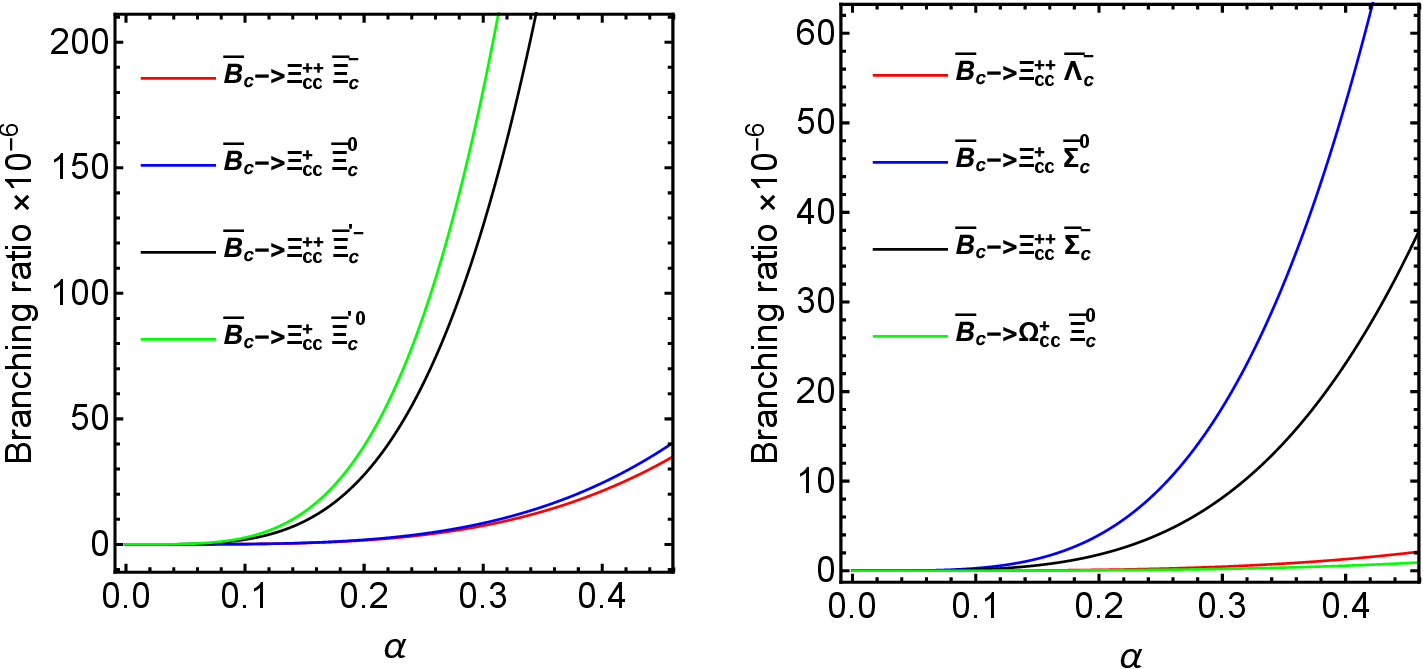}
\caption{Branching ratios respecting to $\alpha$ in $\Lambda$, $\alpha$ varies from 100-300 MeV. Left image contains four Cabibbo allowed processes, right one describes Cabibbo suppressed processes with transition $\bar b\to \bar cc \bar d$.}
\label{fig:alpha}
\end{figure}

Armed with these, we determine the production of doubly charmed baryon from $\overline B_c$ meson, acquiring the numerical number of widths and branching ratios collected into Tab.\,\ref{tab:ratio}, particularly the branching ratio can reach to order $10^{-5}$. In our analysis, we fix the parameter $\alpha$ in $\Lambda$ with a value of 250 MeV. For completeness, the branching ratios respecting to $\alpha$ are studied still in Fig.\,\ref{fig:alpha}. Moreover, we further verify the dominant sub-process of productions $\overline {B}_c\to M_1 M_2 \xrightarrow[]{EP}\overline {\mathcal{B}}_{\bar c} \, \mathcal{B}_{cc}$ ($M_1M_2$ = $J/\psi D,\, \eta_c D,\, J/\psi D^*$ or $\eta D^*$). In regard to the results, we draw several inferences below.
\begin{itemize}
  \item Our calculation shows that the production $\overline B_c  \to \Xi_{cc}^{++} \overline \Xi_{\bar c}^{'-}$ occupies the largest branching ratio among the considering processes in this work, which should be confirmed experimentally. Once verified, it could be an ideal candidate to produce doubly charmed baryon in the future. The dominant contribution of amplitude comes from sub-process $\overline {B}_c\to J/\psi D \xrightarrow[]{\Xi_c}\Xi_{cc}^{++} \overline \Xi_{\bar c}^{'-}$ (branching ratio of the subprocess is $2.1\times 10^{-5}$), and $\overline {B}_c\to \eta_c D^* \xrightarrow[]{\Xi_c}\Xi_{cc}^{++} \overline \Xi_{\bar c}^{'-}$ (branching ratio of the subprocess is $1.1\times 10^{-5})$. Therefore, the major contribution of branching ratio comes from these two sub-processes and their cross ones.
  \item It seems that the leading contributions of strong couplings are subprocesses with both $\mathcal{B}\mathcal{B}V$ or $\mathcal{B}\mathcal{B}P$ vertexes in terms of the processes, i.e, for $\overline B_c \to \Xi_{cc}^{++} \overline \Xi_{\bar c}^{'-}$, the isolated contribution from a pair of $\mathcal{B}\mathcal{B}P$s or $\mathcal{B}\mathcal{B}V$s is triviality, merely at the order of $10^{-8}$.
  \item Different from the first four Cabibbo allowed processes, the last four processes are Cabibbo suppressed. However, the branching ratio are still considerable. Especially, the undiscovered $\Omega_{cc}$ can reach to $10^{-6}$ order.

  \item Tab.\,\ref{tab:ratio} shows the rates of decay widths $\Gamma(\overline B_c \to \Xi_{cc}^{++} \overline \Xi_{\bar c}^{-})/\Gamma(\overline B_c \to \Xi_{cc}^{++} \overline \Lambda_{\bar c}^{-}) \approx 16.4$,\, $\Gamma(\overline B_c \to \Xi_{cc}^{++} \overline \Xi_{\bar c}^{'-})/\Gamma(\overline B_c \to \Xi_{cc}^{++} \overline \Sigma_{\bar c}^{-}) \approx 15.5$, which turn out to be consistent with straightforward symmetry analysis in Eq.\,(\ref{symmetryanalysis}), $|V_{cs}|^2/|V_{cd}|^2 \approx 19.9$. This result exhibits that the SU(3) symmetry is basically maintained in these processes.
\end{itemize}
\begin{table}
  \centering
  \caption{The interpolation parameters of form factors in Eq.\,(\ref{eq:formfactor}), from which $\alpha_1$ is dimensionless and $\alpha_{2,3}$ have dimension GeV$^{-1}$\,\cite{Yao:2021pyf}.}\label{tab:formfactor}
  \begin{tabular}{c c c| c c c c c}\hline
  &\multicolumn{2}{c|}{$\overline B_c\to \eta_c$}& \multicolumn{5}{c}{$\overline B_c\to J/\psi$} \\\hline
                &$F_1(0)$ & $F_0(0)$& $V_0(0)$&$A_0(0)$&$A_1(0)$&$A_2(0)$ & $A_3(0)$ \\
     $\alpha_1$ & 0.632 & 0.632 & 0.836 &0.574 &0.551 &0.561 &0.574\\
     $\alpha_2$ & 0.032 & 0.025 & 0.045 &0.030 &0.018 &0.021 &0.030\\
     $\alpha_3$ & 0.061 & 0.034 & 0.028 &0.055 &0.036 &0.037 &0.055\\
     \hline
   \end{tabular}
\end{table}
\begin{table}
  \centering
  \caption{Branching ratios of doubly charmed baryons from $\overline{B}_c$ ($\overline{B}_c\to M_1 M_2 \xrightarrow[]{EP}\overline {\mathcal{B}}_{\bar c} \, \mathcal{B}_{cc}$) with $\Lambda=m_{\mathcal{B}_{c}}+0.1$ GeV. The exchanged particles (EP) can be $\Lambda_c,\Xi_c, \Sigma_c$. In addition, we show the dominant contribution from the $M_1M_2$. }\label{tab:ratio}
  \begin{tabular}{|c|c|c|c|c|}
     \hline
     Process  & EP& Decay width($\times 10^{-19}$)&Branching ratio($\times 10^{-6}$)& Dominant $M_1M_2$\\\hline
     $\overline B_c  \to \Xi_{cc}^{++} \overline \Xi_{\bar c}^-$ & $\Xi_c$ &4.88 &$3.78$ & $\eta_cD^*$\\
     $\overline B_c \to \Xi_{cc}^+ \overline \Xi_{\bar c}^0 $ & $\Xi_c$ & 5.51&$4.27$ & $ \eta_cD^*$\\
     $\overline B_c  \to \Xi_{cc}^{++} \overline \Xi_{\bar c}^{'-}$ & $\Xi_c^{'}$ &83.1 &$64.4$ & $J/\psi D, \eta_c D^*$\\
     $\overline B_c \to \Xi_{cc}^+ \overline \Xi_{\bar c}^{'0} $ & $\Xi_c^{'}$ & 117.9&$91.4$ & $J/\psi D, \eta_c D^*$\\\hline
     $\overline B_c  \to \Xi_{cc}^{++} \overline \Lambda_{\bar c}^-$ & $\Lambda_c$ &0.30 &$0.23$ & $\eta_cD^*$\\
     $\overline B_c \to \Xi_{cc}^+ \overline \Sigma_{\bar c}^0 $ & $\Sigma_c$ &11.9 &$9.25$ & $J/\psi D,\eta_cD^*$\\
     $\overline B_c  \to \Xi_{cc}^{++} \overline \Sigma_{\bar c}^-$ & $\Sigma_c$ & 5.37&$4.16$ & $J/\psi D, \eta_cD^*$\\
     $\overline B_c \to \Omega_{cc}^+ \overline \Xi_{\bar c}^0 $ & $\Xi_c$ & 0.13&$0.10$ & $\eta_cD^*$\\
     \hline
   \end{tabular}
\end{table}

\section{Conclusions}
The observation of $\Xi_{cc}^{++}$ by the LHCb Collaboration opens a new field of research about the nature of baryons containing two heavy quarks. The doubly charmed baryons can form an SU(3) triplet $\Xi_{cc}^{++}, \Xi_{cc}^{+}$ and $\Omega_{cc}^+$, therefore, after the experimental observation of $\Xi_{cc}^{++}$, it is of prime importance to pursue searches for additional production modes of this particle and the other two. It is worth noting that the charmonium pentaquarks have been discovered from the $B$-meson dacays\,\cite{LHCb:2021chn,LHCb:2022ogu}, this inspired us to search for doubly charmed baryons through meson decays.

In this paper, we have presented a theoretical study on the production of doubly charmed baryon from $\overline B_c$ meson decays. The decay widths and branching ratios are achieved in terms of effective Lagrangian method and collected in Tab.\,\ref{tab:ratio}. We hope these results may provide hints for the exploration of new decay modes in the future.

\section*{Acknowledgments}
Y.X. acknowledges the support by National Natural Science Foundation of China under Grant No. 12005294, and the Fundamental Funds for Key disciplines in Physics with No.2022WLXK05. J.X. is supported by National Natural Science Foundation of China under Grant No. 12105247.
\appendix
\section{couplings}\label{sec:appenda}
The Lagrangian of $\mathcal{\mathcal{B}}\mathcal{B}P$ in SU(3) symmetry can be written as,
\begin{eqnarray}
\mathcal{L}^{SU3}_{\mathcal{B}\mathcal{B}P}=a_1 \overline {\mathcal{B}}_j^i \gamma_5\gamma_{\mu}\partial^{\mu}P_k^j \mathcal{B}_i^k+a_2 \overline {\mathcal{B}}_j^i \gamma_5\gamma_{\mu}\partial^{\mu} P_i^k \mathcal{B}_k^j \,,
\end{eqnarray}
here parameters $a_1=D+F$, $a_2=D-F$ and $\frac{D}{D+F}=\alpha_D=0.64$\,\cite{Adelseck:1990ch}. Expanding the Lagrangian, we take the couplings of $NN\pi$ and $K\Lambda N$ and the rate is
\begin{eqnarray}\label{eq:bbpsu3}
\frac{g_{NN\pi}}{g_{K\Lambda N}}=\frac{\sqrt{3}}{2\alpha_D-3} \,.
\end{eqnarray}
Furthermore, the Lagrangian of $\mathcal{B}\mathcal{B}P$ in SU(4) symmetry is
\begin{eqnarray}
\mathcal{L}^{SU4}_{\mathcal{B}\mathcal{B}P}&=&g_0(b_1 \overline {\mathcal{B}}_{[jk]}^i \gamma_5\gamma_{\mu}\partial^{\mu}P_l^j \mathcal{B}_i^{[kl]}+b_2 \overline {\mathcal{B}}_{[kl]}^i \gamma_5\gamma_{\mu}\partial^{\mu} P_i^j \mathcal{B}_j^{[kl]}) \\
&=& g_0(g_1 \overline {\mathcal{B}}^{\{ij\}k} \gamma_5\gamma_{\mu}\partial^{\mu}P_i^l \mathcal{B}_{\{lj\}k}+g_2 \overline {\mathcal{B}}^{\{ij\}k} \gamma_5\gamma_{\mu}\partial^{\mu} P_i^l \mathcal{B}_{\{lk\}j}) \,.
\end{eqnarray}
Here, we show two equivalent tensor representations of baryon $\mathcal{B}^i_{[\alpha\beta]} \varepsilon^{jk\alpha\beta}$ and $\mathcal{B}^{\{ij\}k}$, $[]$ and $\{\}$ represent symmetric and anti-symmetric indices respectively, $\varepsilon^{jk\alpha\beta}$ is the total antisymmetric tensor. Expanding the Lagrangian
\begin{eqnarray}\label{eq:bbpsu4}
\frac{g_{NN\pi}}{g_{K\Lambda N}}=\frac{b_2}{-\frac{b_1+4b_2}{2\sqrt{3}}}=\frac{-2\sqrt{3}}{\frac{b_1}{b_2}+4}.
\end{eqnarray}
Comparing the results between~(\ref{eq:bbpsu3}) and (\ref{eq:bbpsu4}),
\begin{eqnarray}\label{eq:bbprate}
\frac{b_1}{b_2}=2-4\alpha_D \,.
\end{eqnarray}
It is similar for the case of $\mathcal{B}\mathcal{B}V$ couplings.
Generally, the Lagrangian of $\mathcal{B}\mathcal{B}V$ in SU(3) symmetry can be written as
\begin{eqnarray}
\mathcal{L}^{SU3}_{\mathcal{B}\mathcal{B}V}=c_1 \overline {\mathcal{B}}_j^i \gamma_{\mu} (V^{\mu})_k^j \mathcal{B}_i^k+c_1 \overline {\mathcal{B}}_k^i \gamma_{\mu} (V^{\mu})_i^j \mathcal{B}_j^k+c_2 \overline {\mathcal{B}}_k^i \gamma_{\mu} (V^{\mu})_j^j \mathcal{B}_i^k \,.
\end{eqnarray}
Expanding the Lagrangian and comparing the coupling $\rho NN$ with $K^* \Lambda N$,
\begin{eqnarray}
g_{\rho NN}=-\sqrt{3}g_{K^*\Lambda N} \,.
\end{eqnarray}
The Lagrangian of $BBV$ in SU(4) symmetry is given below,
\begin{eqnarray}
\mathcal{L}^{SU4}_{\mathcal{B}\mathcal{B}V}&=&g'_0(g'_1 \overline {\mathcal{B}}_{[jk]}^i \gamma_{\mu} (V^{\mu})_l^j \mathcal{B}_i^{[kl]}+g'_2 \overline {\mathcal{B}}_{[kl]}^i \gamma_{\mu} (V^{\mu})_i^j \mathcal{B}_j^{[kl]}+g'_3 \overline {\mathcal{B}}_{[kl]}^i \gamma_{\mu} (V^{\mu})_j^j \mathcal{B}_i^{[kl]})\\
&=&g'_0(h_1 \overline {\mathcal{B}}^{\{ij\}k} \gamma_{\mu} (V^{\mu})_i^l \mathcal{B}_{\{lj\}k}+h_2 \overline {\mathcal{B}}^{\{ij\}k} \gamma_{\mu} (V^{\mu})_i^l \mathcal{B}_{\{lk\}j}) \,.
\end{eqnarray}
We again show two equivalent formulas as before. Similarly, one can get
\begin{eqnarray}
\frac{g_{\rho NN}}{g_{K^*\Lambda N}}=\frac{g'_2}{-\frac{g'_1+4g'_2}{2\sqrt{3}}}=\frac{-2\sqrt{3}}{\frac{g'_1}{g'_2}+4} \,,
\end{eqnarray}
finally, we have
\begin{eqnarray}\label{eq:bbvrate}
\frac{g'_1}{g'_2}=-2 \,.
\end{eqnarray}
According to the couplings $g_{NN\pi}=13.5,\, g_{\rho NN}=3.25,\, \kappa_{\rho}=6.1$ as well as the relations in (\ref{eq:bbprate}) and (\ref{eq:bbvrate}), we determine the required couplings,
\begin{eqnarray}
&& g_{\Xi_{c}\Xi_c \eta_c}=\frac{1}{3\sqrt{6}}(b_1-5b_2)=-10.2,\quad g_{\Xi_{c}\Xi^{'}_c \eta_c}=0 \,,\\
&& g_{\Lambda_{c}\Lambda_c \eta_c}=\frac{1}{3\sqrt{6}}(b_1-5b_2)=-10.2,\quad g_{\Lambda_{c}\Sigma_c \eta_c}=0 \,,\\
&& g_{\Sigma_{c}\Sigma_c \eta_c}=\frac{1}{\sqrt{6}}(b_2-b_1)=8.6,\quad g_{\Xi_{c}^{'}\Xi_c^{'} \eta_c}=\frac{1}{\sqrt{6}}(b_2-b_1)=8.6 \,,\\
&& g_{\Xi_{cc}^{++}\Xi_c D_s}=-\frac{1}{\sqrt{3}}(b_1+b_2)=-3.4,\quad g_{\Xi_{cc}^{++}\Xi^{'}_c D_s}=b_2=13.5 \,,\\
&& g_{\Xi_{cc}^{+}\Xi_c D_s}=-\frac{1}{\sqrt{3}}(b_1+b_2)=-3.4,\quad g_{\Xi_{cc}^{+}\Xi^{'}_c D_s}=-b_2=-13.5 \,,\\
&& g_{\Xi_{cc}^{++}\Lambda_c D}=\frac{1}{\sqrt{3}}(b_1+b_2)=3.4,\quad g_{\Xi_{cc}^{++}\Sigma_c D}=b_2=13.5 \,,\\
&& g_{\Xi_{cc}^{+}\Sigma_c D}=-\sqrt{2}b_2=-19.1,\quad g_{\Omega_{cc}^{+}\Xi_c D}=-\frac{1}{\sqrt{3}}(b_1+b_2)=-3.4 \,,\\
&& g_{\Xi_{c}\Xi_c J/\psi}=\frac{1}{3\sqrt{6}}(g_1-5g_2)=-3.10,\quad g_{\Xi_{c}\Xi_c^{'} J/\psi}=0 \,,\\
&& g_{\Lambda_{c}\Lambda_c J/\psi}=\frac{1}{3\sqrt{6}}(g_1-5g_2)=-3.10,\quad g_{\Lambda_{c}\Sigma_c J/\psi}=0 \,,\\
&& g_{\Sigma_{c}\Sigma_c J/\psi}=\frac{g_2-g_1}{\sqrt{6}}=3.98,\quad g_{\Xi_{c}^{'}\Xi_c^{'}J/\psi}=\frac{g_2-g_1}{\sqrt{6}}=3.98 \,,\\
&& g_{\Xi_{cc}^{++}\Xi_c D^*_s}=-\frac{1}{\sqrt{3}}(g_1+g_2)=1.88,\quad g_{\Xi_{cc}^{++}\Xi^{'}_c D^*_s}=g_2=3.25 \,,\\
&& g_{\Xi_{cc}^{+}\Xi_c D^*_s}=-\frac{1}{\sqrt{3}}(g_1+g_2)=1.88,\quad g_{\Xi_{cc}^{+}\Xi^{'}_c D^*_s}=-g_2=-3.25 \,,\\
&& g_{\Xi_{cc}^{++}\Lambda_c D^*}=\frac{1}{\sqrt{3}}(g_1+g_2)=-1.88,\quad g_{\Xi_{cc}^{++}\Sigma_c D^*}=g_2=3.25 \,,\\
&& g_{\Xi_{cc}^{+}\Sigma_c D^*}=-\sqrt{2}g_2=-4.60,\quad g_{\Omega_{cc}^{+}\Xi_c D^*}=-\frac{1}{\sqrt{3}}(g_1+g_2)=1.88 \,.
\end{eqnarray}

  \end{document}